\def\year{2020}\relax
\documentclass[letterpaper]{article} 
\usepackage{aaai20}  
\usepackage{times}  
\usepackage{helvet} 
\usepackage{courier}  
\usepackage[hyphens]{url}  
\usepackage{graphicx} 
\usepackage{graphicx}  
\usepackage{lipsum}
\usepackage{graphicx}
\usepackage{array}
\usepackage{mathrsfs}
\usepackage{amsmath}
\usepackage{amssymb}
\usepackage{amsfonts}
\usepackage{amsthm}
\usepackage{bm}
\usepackage{algorithm}
\usepackage{algorithmic}
\usepackage{color}
\usepackage{tikz}
\usepackage{bbm}
\usepackage{booktabs}
\usepackage{verbatim}
\usepackage{multirow}

\def\eg{e.g.,~}                

\DeclareMathOperator*{\argmin}{\arg\!\min}

\newlength\paramargin
\newlength\figmargin
\newlength\secmargin

\usepackage{color}
\long\def\ignorethis#1{}

\newtheorem{theorem}{Theorem}

\newtheorem{definition}[theorem]{Definition}

\nocopyright
 
 \newcommand{\mbf}[1]{\mathbf{#1}}

\newcommand{\Pa}[1]{\mathsf{PA}_{#1}}
\newcommand{\pa}[1]{\mathsf{pa}_{#1}}

\title{Fairness through Equality of Effort}
\author{Wen Huang, Yongkai Wu, Lu Zhang, Xintao Wu \\ University of Arkansas, Fayetteville, AR, USA \\ \{wenhuang, yw009, lz006, xintaowu\}@uark.edu}

\begin{document}

\maketitle

\begin{abstract}
Fair machine learning is receiving an increasing attention in machine learning fields. Researchers in fair learning have developed correlation or association-based measures such as demographic disparity, mistreatment disparity, calibration,  causal-based measures such as total effect, direct and indirect discrimination, and counterfactual fairness, and fairness notions such as  equality of opportunity and equal odds that consider  both decisions in the training data and decisions made by predictive models.  In this paper, we develop a new causal-based fairness notation, called equality of effort. Different from existing fairness notions which mainly focus on discovering the disparity of decisions between two groups of individuals, the proposed equality of effort notation helps answer questions like to what extend a legitimate variable should change to make a particular individual achieve a certain outcome level and addresses the concerns whether the efforts made to achieve the same outcome level for individuals from the protected group and that from the unprotected group are different. We develop algorithms for determining whether an individual or a group of individuals is discriminated in terms of equality of effort. We also develop an optimization-based method for removing discriminatory effects from the data if discrimination is detected. We conduct empirical evaluations to compare the equality of effort and existing fairness notion and show the effectiveness of our proposed algorithms.    
\end{abstract}

\section{Introduction}

Fair machine learning is receiving an increasing attention in machine learning fields. Discrimination is unfair treatment towards individuals based on the group to which they are perceived to belong. The first endeavor of the research community to achieve fairness is developing correlation or association-based measures, including demographic disparity (e.g., risk difference), mistreatment disparity, calibration, etc. \cite{romei2014multidisciplinary,luong2011k,vzliobaite2011handling,dwork2012fairness,feldman2015},
which mainly focus on discovering the disparity of certain statistical metrics between two groups of individuals. However, as paid increasing attention recently \cite{DBLP:conf/ijcai/ZhangWW17,kilbertus2017avoiding,nabi2018fair}, unlawful discrimination is a causal connection between the challenged decision and a protected characteristic, which cannot be captured by simple correlation or association concepts. To address this limitation, causal-based fairness measures 
have been proposed, including total effect \cite{zhang2018fairness}, direct and indirect discrimination \cite{DBLP:conf/ijcai/ZhangWW17,zhang2018fairness,Chiappa2018Path}, and counterfactual fairness \cite{kusner2017counterfactual,russell2017worlds}. Fairness notions have also been extended to considering both decisions in the training data and decisions made by predictive models, such as equality of opportunity and equal odds \cite{Hardt2016Equality,zafar2017fairness}, and counterfactual direct and indirect error rates \cite{zhang2018equality}. 

In this paper, we develop a new causal-based fairness notation, called equality of effort. 
Consider a dataset with $N$ individuals with attributes $(S, T, \mathbf{X}, Y)$  where $S$ denotes a protected attribute such as {\em gender} with domain values  $\{ s^+, s^-\}$,  $Y$ denotes a decision attribute such as {\em loan} with domain values  $\{ y^+, y^-\}$, $T$ denotes a legitimate attribute such as {\em credit score}, and $\mathbf{X}$ denotes a set of covariates. For a particular applicant $i$ in the dataset with profile $(s^-, t_{i}, \mathbf{x}_{i}, y^-)$, she may ask the counterfactual question, how much her credit score she should improve such that the probability of her loan application approval is above a threshold $\gamma$ (e.g., $80\%$). Informally speaking, our proposed equality of effort notation addresses her concern on whether her future effort (the increase of her credit score) has no difference from male applicants with similar profile $\mathbf{x}$.      

Following Rubin's causal modeling notations, we use $Y_i(t)$ to represent the potential outcome for individual $i$ given a new treatment $T = t$,  $\mathbb{E}[Y_i(t)]$ to denote the individual-level expectation of outcome variable. If $\mathbb{E}[Y_i(t)] \geq \gamma$, we say applicant $i$ tends to receive loan approval with at least probability $\gamma$. We can then calculate or estimate the minimum value of the treatment variable to achieve $\gamma$-level outcome for individual $i$. If the minimum value of individual $i$ is significantly higher than her counterparts (i.e., males with similar characteristics), discrimination exists in terms of effort discrepancy.  
 
Our fairness notation, equality of effort, is different from existing fairness notions,  e.g., statistical disparity, path-specific effects, which mainly focus on the the effect of the sensitive attribute $S$ on the decision attribute $Y$.  Our proposed equality of effort instead focuses on to what extend the treatment variable $T$ should change to make the individual achieve a certain outcome level. This notation addresses the concerns whether the efforts that would need to make to achieve the same outcome level for individuals from the protected group and the efforts from the unprotected group are different. 
 We develop algorithms for determining whether an individual or a group of individuals are discriminated in terms of equality of effort based on three widely used techniques for causal inference,  outcome regression, propensity score weighting, and structural causal modeling. We also develop an optimization-based method for removing discriminatory efforts from biased datasets. We conduct empirical evaluations to compare the equality of effort and existing fairness notions and  evaluation results show the effectiveness of our proposed algorithms.

\section{Preliminaries}
\subsection{Notations}
In this paper, an uppercase denotes a variable, e.g., $S$;
a bold uppercase denotes a set of variables, e.g., $\mbf{X}$; 
a lowercase denotes a value or a set of values of the variables, \eg $s$ and $\mbf{x}$; and  a lowercase with superscript denotes a particular value, \eg $s^+$ and $x^-$.

\subsection{Potential Outcomes Framework}
The potential outcomes framework, also known as Neyman-Rubin  potential  outcomes  or  Rubin  causal  model, has been widely used in many research areas to perform causal inference.  It  refers  to  the  outcomes  one  would  see  under each  treatment  option. Let $Y$ be the outcome variable, $T$ be the binary or multiple valued ordinal treatment variable, and $\mathbf{X}$ be the pre-treatment variables (covariates). 
$Y_i(t)$ represents the potential outcome for individual $i$ given treatment level $T = t$ and $\mathbb{E}[Y_i(t)]$ denotes the individual-level expectation of outcome variable.
The ``fundamental problem of causal inference'' claims that one can never observe all the potential outcomes for any individual \cite{holland1986statistics} and we need to compare potential outcomes and make inference from observed data. 
We  use $\mathbb{E}[Y(t)]$ to denote population-level expectation of outcome variable and $\mathbb{E}[Y_{\diamond}(t)]$ to denote the conditional expectation of outcome variable within certain sub-population $\diamond$.

Traditional causal inference focuses on estimating the potential outcome and treatment effect given the information of treatment variable and pre-treatment variables~\cite{burgette2017propensity} . For example, the average treatment effect, $ATE = \mathbb{E}[Y(t')-Y(t)]$ answers the question of how, on average, the outcome of interest $Y$ would change if everyone in the population of interest had been assigned to a particular treatment $t'$ relative to if they had all received another treatment $t$. The average treatment effect on the treated, $ATT = \mathbb{E}[Y(t')-Y(t)|T=t]$ is about how the average outcome would change if everyone who received one particular treatment $t$ had instead received another treatment $t'$. Under the potential outcomes framework, the outcome function usually has two forms: the regression form and the probability factorization form. Under certain assumptions we can represent the whole inverse process and derive corresponding inverse outcome function.
 
\subsection{Propensity Score Method}
One major challenge in causal inference is the presence of confounding variables. A confounder is the covariate  that affects treatment variable and outcome variable simultaneously. Under the unconfoundedness assumption (no hidden confounders), propensity score method, as a widely used approach to achieve causal inferences from observational data, can  reduce  the  selection  bias  caused  by confounders. 

\begin{definition}[Propensity Score]\label{def:ps}
For a binary treatment variable, propensity score is the conditional probability of receiving treatment $T$ given the pre-treatment variables $\mathbf{X}$,
\[ e(\mathbf{x}) = Pr( T = 1| \mathbf{X}=\mathbf{x}) \]
\end{definition}
The estimation of propensity scores requires the model or functional form of $e(\cdot)$ and the variables to include in $\mathbf{X}$. 
Let $e(i)$ denote the propensity score for individual $i$, for binary valued groups, the propensity score is estimated by logistic regression:
	\[ logit(e(i)) = \beta_0 + \beta_1 x_1 +  ... + \beta_k x_k, \]
where $x_1, ..., x_k$ are values of the selected covariates and $\beta_1,...,\beta_k$ are regression coefficients.	

If correctly estimated, the reciprocal of propensity score can be used as the weight for each individual such that the distribution of the group under treatment 1 and that  under treatment 0 becomes identical. In other words, $Pr(Y_i(1)) =\omega_i \cdot Pr(Y_i(0))$ where $\omega_i = e(i)^{-1}$ is the estimator of the inverse propensity score for individual $i$.  \cite{rosenbaum1983central} showed that conditional on the propensity score, all observed covariates are independent of treatment assignment, and they will not confound estimated treatment effects.  
Hence after weighting procedure, a pseudo-balanced population can be built in which the imbalance caused by measured covariates between the treatment groups has been eliminated. The average potential outcome can thus be estimated by some standard estimators. For example, one unbiased estimator of the population-level $ATE$  can be written as: $\frac{1}{N_1} \sum_{i\in N}\mathbbm{1}_{T_i = 1} w_i y_i - \frac{1}{N_2} \sum_{i \in N}\mathbbm{1}_{T_i = 0} w_i y_i$ where  $N_1 = \sum_{i\in N} \mathbbm{1}_{T_i = 1}$ and $N_2 = \sum_{i\in N}  \mathbbm{1}_{T_i = 0}$.  

\section{Fairness through Equal Effort}

We assume a population with attributes $(S, T, \mathbf{X}, Y)$  where $S$ denotes a protected attribute with domain values  $\{ s^+, s^-\}$,  $Y$ denotes a decision attribute with domain values  $\{ y^+, y^-\}$, $T$ denotes a legitimate attribute, and $\mathbf{X}$ denotes a set of covariates. Without losing generality, we assume  there is only one binary protected attribute, one binary decision attribute, and one ordered multi-category legitimate attribute. In this paper, we simply use the change of $T$ as the effort needed to achieve a certain level of outcome and do not consider the real monetary or resource cost behind that change.      

\subsection{Equality of Effort at the Individual Level}
For an individual $i$ in the dataset with profile $(s_{i}, t_{i}, \mathbf{x}_{i}, y_{i})$,  we want to figure out what is the minimal change on treatment variable $T$ to achieve a certain outcome level based on observational data. If the minimal change for individual $i$ has no difference from that of counterparts (individuals with similar profiles except the sensitive attribute), we say  individual $i$ achieves fairness in terms of equality of effort.   

Formally, we use $Y_i(t)$ to represent the potential outcome for individual $i$ given a new or counterfactual treatment $T = t$. We use $\mathbb{E}[Y_i(t)]$ to denote the individual-level expectation of outcome variable where $\mathbb{E}[\cdot]$ is the expectation operator from probability theory.  When $\mathbb{E}[Y_i(t)]$ is larger than a predefined threshold $\gamma$, we say individual $i$  would receive a positive decision with probability $\gamma$.

\begin{definition}[$\gamma$-Minimum Effort]
\label{def:min_effort}
 For individual $i$ with value $(s_{i}, t_{i}, \mathbf{x}_{i}, y_{i})$, the minimum value of the treatment variable to achieve $\gamma$-level outcome is defined as:
 \[
   \Psi_i (\gamma) = \argmin_{t\in T} \big\{ \mathbb{E}[Y_i(t)] \geq \gamma)     \} 
 \]
and the minimum effort to achieve  $\gamma$-level outcome is $\Psi_i (\gamma)- t_{i}$.  
\end{definition}

However $Y_i(t)$ cannot be directly observed and we have to derive its estimate from samples with similar characteristics. 
We design an estimation procedure based on the idea of situation testing, which is one normal practice of determining whether an individual is discriminated. How to select variables for finding similar individuals has been studied in situation testing based individual discrimination discovery \cite{zhang2016b}. The proposed idea there was to first construct a causal graph for all variables and then select variables that are the parents of the decision. Their work is also applicable to our equal effort definition. We first find a subset of users, denoted as $I$, each of whom has the same (or similar) characteristics ($\mathbf{x}$ and $t$) as individual $i$. We denote $I^+$ ($I^-$) the subgroup of users in $I$ with the sensitive attribute value $s^+$ ($s^-$).  Similarly, $\mathbb{E}[Y_{I^+}(t)]$ denotes the expected outcome under treatment $t$ for the subgroup $I^+$.   
The minimal effort needed to achieve $\gamma$ level of outcome variable within the subgroup $I^+$ is then defined as: 
     \[ \Psi_{I^+}(\gamma) =  \argmin_{t\in T} \{\mathbb{E}[Y_{I^+}(t)] \geq \gamma \}.\]

\begin{definition}[$\gamma$-Equal Effort Fairness at the Individual Level]
\label{def:equ_effort_i}  
For a certain outcome level $\gamma$, we define equality of effort for individual $i$ if  
\[
\Psi_{I^+}(\gamma) = \Psi_{I^-}(\gamma).
\]
The difference $\delta_i(\gamma) = \Psi_{I^+}(\gamma) - \Psi_{I^-}(\gamma)$ measures the effort discrepancy at the individual level. 
\end{definition}

\subsection{Equality of Effort at the Group or System Level}
In addition to the task of checking individual level discrimination, we also want to check whether discrimination exists at the group or system level. System-level discrimination deals with the average discrimination across the whole system, e.g., all applicants to a university,  and group-level discrimination deals with discrimination that occurs in one particular subgroup, e.g., the applicants applying for a particular major. Existing works \cite{vzliobaite2011handling,DBLP:conf/ijcai/ZhangWW17} apply demographic disparity metrics (e.g., risk difference) or causal effect (e.g., direct and indirect causal discrimination) on the whole dataset (the subset of data) to determine the system-level (group-level) discrimination. Similarly, we may want to check whether there are effort discrepancies at the group or system level. 

We denote $D$ as the whole dataset,  and $D^+$ ($D^-$) as the subset with the sensitive attribute value $s^+$ ($s^-$).  We define the minimum value of treatment variable to achieve a certain outcome level $\gamma$ for $D^*$  as:
 \[
   \Psi_{D^*} (\gamma) = \argmin_{t\in T} \big\{ \mathbb{E}[Y_{D^*}(t)] \geq \gamma)     \big\}. 
 \]

\begin{definition}[$\gamma$-Equality of Effort at the System Level]
\label{def:equ_effort_s} 
  For a certain outcome level $\gamma$, equality of effort between two sensitive attributes $s^+$ and $s^-$ is achieved if
  \[
      \Psi_{D^+}(\gamma)  =   \Psi_{D^-}(\gamma).
  \]
  The difference $\delta_D(\gamma) = \Psi_{D^+}(\gamma) - \Psi_{D^-}(\gamma)$ measures the effort discrepancy at the system level.
 \end{definition}

Definition \ref{def:equ_effort_s} can be straightforwardly adapted to the group level. Given two compared groups, their distributions in terms of certain attributes (e.g., outstanding debt) could be different. The simply use of our group equal-effort fairness may not be appropriate. In this case, we could apply the path-specific effect/mediator analysis \cite{DBLP:conf/ijcai/ZhangWW17,nabi2018fair} to separate and measure different causal effects e.g., direct discrimination, indirect discrimination, and explainable effects.

\subsection{Comparison with Other Fairness Metrics}

\begin{table*}\small
\centering
\begin{tabular}{lll} \hline
 Notation & References & Formula  \\ \hline
    Demographic parity  &\cite{verma2018fairness}  & $P(y^+|s^+) - P(y^+|s^-)$ \\
    Conditional parity  & \cite{verma2018fairness} & $P(y^+|s^+,\mathbf{o}) - P(y^+|s^-,\mathbf{o})$ \\ 
	Total causal discrimination & \cite{DBLP:conf/ijcai/ZhangWW17,zhang2018fairness} & $\mathbb{E}[Y(s^+)] - \mathbb{E}[Y(s^-)]$ \\                                    	Path-specific causal discrimination & \cite{DBLP:conf/ijcai/ZhangWW17,nabi2018fair} & $\mathbb{E}[Y(s^+)|\pi] - \mathbb{E}[Y(s^-)|\pi]$ \\	
	Counterfactual fairness & \cite{kusner2017counterfactual}   & $\mathbb{E}[Y_{\mathbf{o}}(s^+)] - \mathbb{E}[Y_{\mathbf{o}}(s^-)]$ \\
	PC Fairness & \cite{DBLP:journals/corr/abs-1910-12586}    & $\mathbb{E}[Y_{\mathbf{o}}(s^+)|\pi] - \mathbb{E}[Y_{\mathbf{o}}(s^-)|\pi$] \\\hline
    Equality of opportunity                 & \cite{Hardt2016Equality,zafar2017fairness}  & $P(\hat{Y}=y^+|s^+,y^+) - P(\hat{Y}=y^+|s^-, y^+)$ \\
    Calibration                             & \cite{Hardt2016Equality,zafar2017fairness}  & $P(y^+|s^+,\hat{Y}=y^+) - P(y^+|s^-, \hat{Y}=y^+)$ \\ \hline          
\end{tabular}
\caption{Formula of previous fairness notions}
\label{tab:difference}
\end{table*}

Many different fairness metrics have been proposed to measure fairness of data and machine learning algorithms. Classic metrics include individual fairness, demographic parity, equality of opportunity, calibration, causal fairness, and counterfactual causal fairness. Refer to a recent survey \cite{verma2018fairness}. We show in Table \ref{tab:difference}  the formula of previous representative fairness metrics to compare with our equality of effort notion.   For example, demographic imparity requires that $P(y^+|s^+)=P(y^+|s^-)$ and similarly conditional demographic imparity requires $P(y^+|s^+, \mathbf{o})=P(y^+|s^-,\mathbf{o})$ where $\mathbf{o}$ is the values of a specified variable set $\mathbf{O}$. Basically they require that a decision be independent of the protected attribute conditional or unconditional on some other variables.   
For causal based fairness notions, the total causal discrimination is based on the average causal effect of $S$ on $Y$ and is defined as $\mathbb{E}[Y(s^+)] - \mathbb{E}[Y(s^-)]$,  which represents the expected change of outcome $Y$ when $S$ of all individuals changes from $s^-$ to $s^+$. Different from the total causal discrimination that measures the causal effect transmitted along all the causal paths from $S$ to $Y$ in the causal graph, the path-specific causal discrimination is based on the causal effect that is transmitted along some specific paths $\pi$ from $S$ to $Y$, e.g., direct causal discrimination when $\pi$ is the direct path from $S$ to $Y$, and indirect causal discrimination  when $\pi$ is all paths from $S$ to $Y$ through redlining attribute $T$. Counterfactual fairness requires $\mathbb{E}[Y_{\mathbf{o}}(s^+)] - \mathbb{E}[Y_{\mathbf{o}}(s^-)]$, which means that a decision is fair towards an individual if it is the same in the actual world and  a counterfactual world where the individual belonged to a different demographic group. Most recently, \cite{DBLP:journals/corr/abs-1910-12586} developed a unified definition, path-specific counterfactual fairness (PC Fairness), that covers previous causality-based fairness notations. Different from demographic parity and causal based fairness notions, our proposed equality of effort considers to what extend the legitimate variable $T$ should change to achieve a certain outcome level and whether the minimum effort made for individuals from the protected group and that from the unprotected group are the same. 

When considering discrimination from the perspective of supervised learning, the equality of opportunity is based on the actual outcome $Y$ and the predicted outcome $\hat{Y}$,  requiring $P(\hat{Y}=y^+|s^+,y^+) = P(\hat{Y}=y^+|s^-, y^+)$. Basically it  means the decision model should not mistakenly predict examples with $y^+$ as $\hat{Y} = y^-$ at a higher rate for one group than another. In other words, a predictor $\hat{Y}$ satisfies equalized opportunity with respect to protected attribute $S$ and outcome $Y$ if $\hat{Y}$ and $S$ are independent conditional on $Y$. Similarly the calibration considers the fraction of correct positive predictions and requires $P(y^+|s^+,\hat{Y}=y^+) = P(y^+|s^-, \hat{Y}=y^+)$. Our proposed equality of opportunity does not consider the model predictions and instead focuses on the effort, i.e., the minimum change of $T$ to achieve a certain outcome level $Y$, based on the causal framework.     

We noticed a parallel work \cite{DBLP:journals/corr/abs-1903-01209} that developed an effort-based measure of fairness and formulated effort unfairness as the inequality in the amount of effort required for members from disadvantage group and advantaged group. However, their work focused on characterizing the long-term impact of algorithmic policies on reshaping the underlying population based on the psychological literature on social learning and the economic literature on equality of opportunity. Our work is based on counterfactual causal inference and develops an optimization-based framework for removing discriminatory effort unfairness from the static data if discrimination is detected. 

\section{Calculating Average Effort Discrepancy}

In real-world applications, we often have multiple values of $\gamma$ used in decision making. 
We use the average effort discrepancy over all values of $\gamma$ as the measure of equality of effort in this scenario. If $\gamma$ has a set of discrete values, then the average is computed by the mean of all effort discrepancies. If $\gamma$ is a continuous variable, then the average is defined as the integration over the range of $\gamma$.

\begin{definition}[Average Effort Discrepancy (AED)]
\label{def:aed}
If $\gamma \in \Gamma$ where $\Gamma$ denotes the effort level value set of the expectation of outcome variable, then the average effort discrepancy is defined as 
\begin{equation} \label{eq:aed_d}
AED = \frac{1}{|\Gamma|}\sum_{\gamma \in \Gamma} \delta(\gamma).
\end{equation}
If $\gamma$ is a continuous variable in a range $[\gamma_{1},\gamma_{2}]$, then the average effort discrepancy is defined as 
\begin{equation} \label{eq:aed_c}
AED =  \frac{1}{\gamma_2 - \gamma_1}  \int_{\gamma_1}^{\gamma_2} \delta(\gamma) d\gamma.
\end{equation}
  \end{definition}

To calculate the AED, we need to first compute the expected outcome $\mathbb{E}[Y_{I^{*}}(t)]$ or $\mathbb{E}[Y_{D^{*}}(t)]$, and then compute the minimum effort. In the following, we develop a general calculating method assuming the monotonicity and invertibility for $\mathbb{E}[Y_{D^{*}}(t)]$. Then, we consider three widely used techniques for causal inference: outcome regression and propensity score weighting from Rubin's framework, and structural causal analysis from Pearl's framework. We compute the AED for each of the techniques.

\begin{algorithm}[h]
\caption{Discrimination detection through equal effort}
\textbf{Input} Dataset $D$, Threshold $\tau$ \\
\textbf{Output} Result 
\begin{algorithmic}[1]
\STATE For each subset $D^* \in \{D^+,D^-\}$, identify expected outcome $f_{D^*}(t)=\mathbb{E}[Y_{D^*}(t)]$
\IF{$f_{D^*}(t)$ is continuous, monotonous and invertible}
    \STATE Calculate $AED$ according to Eq.~\eqref{eq:aed_close}
\ELSE 
    \STATE Identify inverse function $f^{-1}_{D^*}(\gamma)$
    \IF{$f^{-1}_{D^*}(\gamma)$ has a closed form}
        \FOR{each $\gamma$}
            \STATE Find the minimum value of $t$ such that $t \geq f^{-1}_{D^*}(\gamma)$
            \STATE Calculate effort discrepancy $\delta_D(\gamma)$
        \ENDFOR
    \ELSE
        \FOR{each treatment level $t$}
            \STATE Use appropriate causal inference method to estimate $\hat{\mathbb{E}}[Y_{D^*}(t)]$
        \ENDFOR
        \FOR{each $\gamma$}
            \STATE Numerically find the minimum value of $t$ such that $\hat{\mathbb{E}}[Y_{D^*}(t)]\geq \gamma$
            \STATE Calculate effort discrepancy $\delta_D(\gamma)$
        \ENDFOR
        \STATE Calculate $AED$ following Definition \ref{def:aed}
    \ENDIF
\ENDIF

\IF { $|AED|\geq \tau$ }
     \STATE Result = True  
\ELSE
     \STATE Result = False
\ENDIF

\end{algorithmic}
\label{alg:ipw}
\end{algorithm}

Algorithm \ref{alg:ipw} shows the pseudocode of our algorithm for computing the AED and making the judge of discrimination through equal effort. Lines 2-3 deal with the situation where $f_{D^*}(t) = \mathbb{E}[Y_{D^{*}}(t)]$ is a continuous, monotonous and invertible function of $t$, and AED can be directly computed through an integration over $f_{D^*}(t)$ given in the next subsection. If the assumptions are not satisfied, lines 6-10 handle the situation where the closed-form of inverse function $f^{-1}_{D^*}(\gamma)$ can be derived; and lines 12-19 handle the situation otherwise.

\subsection{General Method under Monotonicity and Invertibility Assumption}
As discussed in the previous section, $\mathbb{E}[Y_{D^{+}}(t)]$ and $\mathbb{E}[Y_{D^{-}}(t)]$ denote the expectations of outcome variable for groups $D^{+}$ and $D^{-}$. We can treat them as functions of $t$, denoted as $f_{D^{+}}(t)$ and $f_{D^{-}}(t)$. Under the assumptions of being monotonically increasing and invertible, inequality $\mathbb{E}[Y_{D^{+}}(t)]\geq \gamma$ can be expressed as $f_{D^{+}}(t)\geq \gamma$, which leads to $t\geq f_{D^{+}}^{-1}(\gamma)$, where $f_{D^{+}}^{-1}(\cdot)$ is the inverse function of $f_{D^{+}}(\cdot)$. As a result, we directly obtain that $\Psi_{D^+} (\gamma) = f_{D^{+}}^{-1}(\gamma)$, and similarly $\Psi_{D^-} (\gamma) = f_{D^{-}}^{-1}(\gamma)$.

If the closed forms of $f_{D^{+}}^{-1}(\cdot)$ and $f_{D^{-}}^{-1}(\cdot)$ can be derived, then the AED can be easily computed; otherwise its calculation is not straightforward. However, when $\gamma$ is a continuous variable, then we don't need to derive the closed form of the inverse functions to compute the AED, but only require the integration of $f_{D^{+}}(\cdot)$ and $f_{D^{-}}(\cdot)$ to be tractable. This is because based on the Laisant's theorem we have
\[ \int_{\gamma_{1}}^{\gamma_{2}} f^{-1}_{D^{+}} (\gamma) d\gamma = \gamma_{2}t_{2}^{+}-\gamma_{1}t_{1}^{+} - \int_{t_{1}^{+}}^{t_{2}^{+}} f_{D^{+}} ( \gamma) d \gamma, \]
where $t_{1}^{+} = f_{D^{+}}^{-1}(\gamma_{1})$ and $t_{2}^{+} = f_{D^{+}}^{-1}(\gamma_{2})$. In practice, $t_{1}^{+}$ and $t_{2}^{+}$ can be estimated using numerical methods. As a result, the AED is given by
\begin{equation}\label{eq:aed_close}
    (t_{2}^{+}-t_{2}^{-})\gamma_{2} - (t_{1}^{+}-t_{1}^{-})\gamma_{1} - \Big( \!\! \int_{t_{1}^{+}}^{t_{2}^{+}} \!\! f_{D^+} (\gamma)d \gamma - \int_{t_{1}^{-}}^{t_{2}^{-}} \!\! f_{D^-} (\gamma)d \gamma\Big).
\end{equation}

\subsection{Outcome Regression}
Outcome regression is one straightforward method to conduct causal inference. In this approach, a model is posited for the outcome variable as a function of the treatment variable and the covariates. The basic outcome regression model is the linear regression of the form:
	\[ \mathbb{E}[Y|T,\mathbf{X}] = \beta_0 + \beta_1 T +  \bm{\beta}_2 \bm{X} + \bm{\beta}_3\bm{X}T, \]
where $\beta_0, \beta_1$ are regression coefficients, $\bm{\beta_2}$ and $\bm{\beta_3}$ are the coefficient vectors with the same length as $\bm{X}$. All the parameters can be estimated by least squares method.	

One advantage of outcome regression  is it can help us directly calculate the relative treatment value given a certain expected outcome level.
Suppose the regression model is correctly specified, the expected outcome of any subset $D^{*}$ is given by
\[ \mathbb{E}[Y_{D^{*}}(t)] = \frac{1}{|D^{*}|}\sum_{i\in D^{*}}(\beta_0 + \beta_1 t +  \bm{\beta}_2 \bm{x}_{i} + \bm{\beta}_3\bm{x}_{i}t). \]
Thus, the minimum value of the treatment variable to achieve $\gamma$-level outcome, i.e., $\Psi_{D^{*}} (\gamma)$, can be expressed as:
\begin{equation} \label{eq:psi_r}
\argmin_{t\in T} \big\{ \mathbb{E}[Y_{D^{*}}(t)] \geq \gamma)   =  \frac{ \gamma-\frac{1}{|D^{*}|}\sum_{i\in D^{*}}(\beta_{0}+\bm{\beta}_{2}) }{ \frac{1}{|D^{*}|}\sum_{i\in D^{*}}(\beta_{1}+\bm{\beta}_{3}) }.
\end{equation}

\subsection{Propensity Score Weighting}
Another widely used branch of  causal  inference is based on weighting and a typical method is the inverse propensity score weighting. In our context, the treatment variable is a multiple valued ordinal variable, we apply generalized propensity score \cite{imbens2000role} to estimate the weights.

\begin{definition}[Generalized Propensity Score] 
The generalized propensity score for individual $i$ is the conditional probability of receiving a particular level of the treatment given the pre-treatment variables:
\[ r(t,\mathbf{x_i}) = Pr( T = t| \mathbf{X_i}=\mathbf{x_i}). \]
\end{definition}

The weighted mean of the potential outcomes for those who received the treatment $t$ had they received another treatment $t'$ can be consistently estimated by
	\[ \hat{\mathbb{E}}[Y(t')|t]  = \frac{\sum_{i\in N}\mathbbm{1}_{T_i = t'} Y_i \omega_i(t,t') }{\sum_{i\in N}\mathbbm{1}_{T_i = t'} \omega_i(t,t') },  \]
where
	\[ \omega_i(t,t') = \frac{r(t,\mathbf{x_i})}{r(t',\mathbf{x_i})}. \]
Following the above method,  we can get a table showing estimation values of the expected outcome under all treatment pair combinations  $(t,t')$. Thus, the minimum treatment value to achieve $\hat{\mathbb{E}}[Y(t')|t] \geq \gamma $ can be determined by comparing the results in that table.

\subsection{Structural Causal Model}
The structural causal model describes the causal mechanisms of a system as a set of structural equations. For ease of representation, each causal model can be illustrated by a directed acyclic graph called the causal graph, where each node represents a variable and each edge represents the direct causal relationship specified by the causal model. In addition, each node $V$ is associated with a conditional probability distribution $P(v|\pa{V})$ where $\pa{V}$ is the realization of a set of variables $\Pa{V}$ called the parents of $V$. The treatment is modeled using the intervention, which forces the treatment variable $T$ to take certain value $t$, formally denoted by $do(T=t)$ or $do(t)$. The potential outcome of variable $Y$ under intervention $do(t)$ is denoted as $Y_{t}$. The distribution of $Y_{t}$, also referred to as the post-intervention distribution of $Y$ under $do(t)$, is denoted as $P(Y_t)$. Facilitated by the intervention, the expected outcome $\mathbb{E}[Y_{D^{*}}(t)]$ can be measured by the counterfactual quantity $\mathbb{E}[Y_{t}|\mathbf{z}^{*}]$, where $\mathbf{z}^{*}$ represents attribute values that form the subgroup $D^{*}$. The counterfactual quantity measures the expected outcome of $Y$ assuming that the intervention is performed on the subgroup of individuals only. According to \cite{pearl2009causality}, if attributes $\mathbf{Z}$ are non-descendant of $T$ in the causal graph, then $P(Y_{t}|\mathbf{z}^{*})$ can be computed from observational data as
\begin{equation*}
    \frac{ \sum_{\mathbf{X}\backslash \mathbf{Z}} \prod_{V\in \{Y,S,\mathbf{X}\}}  P(v|\pa{V})_{\delta_{T=t}}}{P(\mathbf{z}^{*})},
\end{equation*}
where $\delta_{T=t}$ means assigning $T$ involved in all probabilities with the corresponding value $t$.

If the inverse function of $\mathbb{E}[Y_{t}|\mathbf{z}^{*}]$ can be derived, then we follow lines 6-10 in Algorithm \ref{alg:ipw} to compute AED; otherwise, we follow lines 12-19 to compute AED.

\section{Achieving Equal Effort}
When our discrimination detection algorithm shows that a dataset does not satisfy the equal effort requirement, then we may want to remove the discriminatory effects from the dataset before it is used for any predictive analysis, i.e., training a decision model. In this section, we develop a method for generating a new dataset which is close to the original dataset and also satisfies equal effort. Our removal method is based on the  use of outcome regression to estimate the potential outcome, but it can be easily extended to any method where the closed form of $\Psi (\gamma)$ can be derived. The general idea is to derive a new outcome regression model satisfying the equal effort constraints. Then, for each individual in the original dataset, we randomly generate a new value $\tilde{Y}$ based on the expectation computed from the fair outcome regression model. 

Specifically, we consider two outcome regression models for subsets $D^{+}$ and $D^{-}$ respectively, given by
\[ \mathbb{E}[Y_{D^{+}}|T,\mathbf{X}] =  \beta_{0}^{+} + \beta_{1}^{+} T +  \bm{\beta}_{2}^{+} \bm{X} + \bm{\beta}_{3}^{+}\bm{X}T, \]
\[ \mathbb{E}[Y_{D^{-}}|T,\mathbf{X}] =  \beta_{0}^{-} + \beta_{1}^{-} T +  \bm{\beta}_{2}^{-} \bm{X} + \bm{\beta}_{3}^{-}\bm{X}T. \]
Then, as shown by Eq.~\eqref{eq:psi_r}, the minimum effort for subgroup $D^{+}$ (and similarly for subgroup $D^{-}$) is given by
\[ \Psi_{D^{+}}(\gamma) = \frac{ \gamma-\frac{1}{|D^{+}|}\sum_{i\in D^{+}}(\beta_{0}^{+}+\bm{\beta}_{2}^{+}) }{ \frac{1}{|D^{+}|}\sum_{i\in D^{+}}(\beta_{1}^{+}+\bm{\beta}_{3}^{+}) }. \]
As a result, the AED according to either Eq.~\eqref{eq:aed_d} or \eqref{eq:aed_c} is given by 
\[ \frac{ \bar{\gamma}-\frac{1}{|D^{+}|}\sum_{i\in D^{+}}(\beta_{0}^{+}+\bm{\beta}_{2}^{+}) }{ \frac{1}{|D^{+}|}\sum_{i\in D^{+}}(\beta_{1}^{+}+\bm{\beta}_{3}^{+}) } - \frac{ \bar{\gamma}-\frac{1}{|D^{-}|}\sum_{i\in D^{-}}(\beta_{0}^{-}+\bm{\beta}_{2}^{-}) }{ \frac{1}{|D^{-}|}\sum_{i\in D^{-}}(\beta_{1}^{-}+\bm{\beta}_{3}^{-}) }, \]
where $\bar{\gamma}$ equals $\frac{1}{|\Gamma|}\sum_{\gamma \in \Gamma}\gamma$ if discrete and $\frac{\gamma_{2}^{2}-\gamma_{1}^{2}}{2}$ if continuous. We want the AED to approach zero. After adding the penalty term for the AED, the objective function becomes
\[ \argmin_{\beta} \sum_{i\in D^{+},D^{-}} \left( \mathbb{E}[Y_{D^{*}}|t_{i},\mathbf{x}_{i}]-y_{i} \right)^{2}+\lambda \cdot AED^2 \]
where $D^{*}=$ $D^{+}$ or $D^{-}$ and $\lambda$ is the parameter for balancing the two objectives.

Finally, for each individual $i$ in the dataset with profile $(s_{i}, t_{i}, \mathbf{x}_{i}, y_{i})$, we first compute his expected value of $Y$ using the fair outcome regression model, i.e., $\mathbb{E}[Y_{D^{*}}|t_{i},\mathbf{x}_{i}]$, where $D^{*}=$ $D^{+}$ or $D^{-}$ depending on the value of $s_{i}$. Then, we randomly assign $0$ or $1$ to the new value $\tilde{y}_{i}$ based on the probability given by $\mathbb{E}[Y_{D^{*}}|t_{i},\mathbf{x}_{i}]$. The generated data then satisfies the equal effort requirement.

\begin{table}[]
	\begin{tabular}{|l|l|}
		\hline
		Category & Original Values                            \\ \hline
		$0$              & Preschool, 1st-4th, 5th-6th                \\ \hline
		$1$              & 7th-8th, 9th, 10th                         \\ \hline
		$2$              & 11th, 12th, HS-grad                        \\ \hline
		$3$              & Some-college, Assoc-voc, Assoc-acdm        \\ \hline
		$4$              & Bachelors, Masters, Prof-school, Doctorate \\ \hline
	\end{tabular}
\caption{Preprocessing \textit{education}.}
\label{tab:edu}
\end{table}

\section{Experiments}
\begin{figure}
  \centering\includegraphics[width=0.7\linewidth]{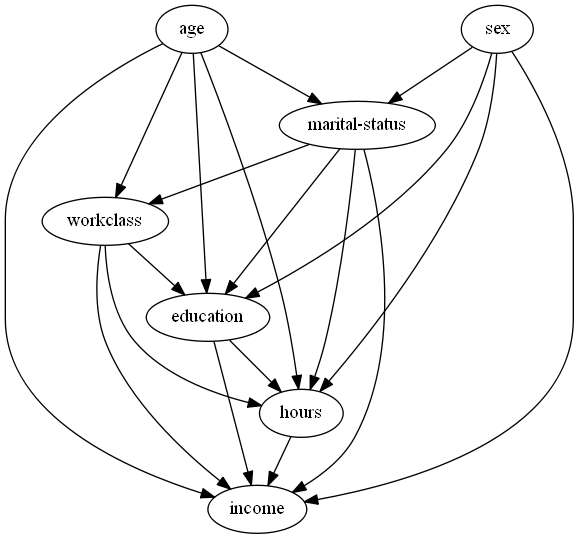}
  \caption{Causal Graph for the Adult Dataset.}
  \label{fig:adult}
\end{figure}

We evaluate our discrimination detection and removal algorithms based on the proposed equality of effort on the UCI Adult dataset \cite{lichman2013}. The Adult dataset contains $65,123$ records with $14$ attributes. 
We select $7$ attributes, \textit{sex},  \textit{age},  \textit{marital status},  \textit{workclass},  \textit{education},  \textit{hours}, and \textit{income} in our experiments.  We consider \textit{income} as the outcome, \textit{education} as the treatment attribute, and \textit{sex} as the protected attribute. Due to the sparse data issue, we binarize the domain of \textit{age},  \textit{marital status},  \textit{workclass},  and \textit{hours} into two classes. We also categorize 16 values of \textit{education} into five levels, as shown in Table~\ref{tab:edu}.

In our experiments, we calculate the minimum effort based on three methods, outcome regression (\textit{Regression}), propensity score weighting (\textit{Weighting}), and structural causal model inference (\textit{SCM}). For \textit{Weighting}, we implement the propensity score weighting for multiple treatments by following the work of  \cite{mccaffrey2013tutorial} and \cite{burgette2017propensity}. For \textit{SCM}, we follow the settings of \cite{DBLP:conf/ijcai/ZhangWW17} and use three tiers for causal graph learning: \textit{sex}, \textit{age} in Tier 1, \textit{marital-status}, \textit{education}, \textit{workclass}, and \textit{hours} in Tier 2, and \textit{income} in Tier 3.  The causal graph is constructed and presented by utilizing the open-source software TETRAD \cite{scheines1998tetrad}. We employ the original PC algorithm \cite{spirtes2000causation} and set the significance threshold 0.01 for conditional independence setting in causal graph construction. Figure \ref{fig:adult} shows the built causal graph. We apply the nonparametric inference of the structural causal model by following the work of \cite{zhang2017a}.  In discrimination removal, the quadratic programming is solved using PyTorch \cite{paszke2017automatic}. 

\subsection{Discrimination Discovery}

\subsubsection{Checking Equal Effort at the System Level} Table~\ref{tab:system-level} shows the comparison results of the expectations of the potential outcome for males ($\mathbb{E}[Y_{D^+}(t)]$) and that for females  ($\mathbb{E}[Y_{D^-}(t)]$) in Adult. We calculate the expectation of the potential outcomes using three methods, \textit{Weighting}, \textit{Regression}, and \textit{SCM}, and vary the treatment variable \textit{education} from $0$ to $4$. As shown in  Table~\ref{tab:system-level}, the expectations of potential outcome for males are significantly higher than the corresponding values for females, indicating large effort discrepancy exists in Adult. For example,  $\mathbb{E}[Y_{D^+}(t)] = 0.498$ and   $\mathbb{E}[Y_{D^-}(t)] = 0.221$ when $t=2$ based on \textit{SCM}. If we set $\gamma = 0.7$, the minimum values of treatment variable (\textit{education}) to achieve $\gamma$-level outcome are $3$ for males (with the expectation of the potential outcome $0.741$) and $4$ for females (with the expectation of the potential outcome $0.706$). The effort discrepancy between females and  males is $1$, which indicates the existence of significant discrimination in terms of equal effort fairness. We would like to point out that the expectations of potential outcome calculated from three methods are generally consistent as shown in Table~\ref{tab:system-level}. However, each calculation method has its own applicable assumptions and may not achieve reliable results when those assumptions are not met. There are extensive researches on the applicability of those causal inference methods (e.g., refer to \cite{pearl2009causality}), which are out of the scope of this work.

\begin{table}[]
\centering
\scalebox{0.7}{
\begin{tabular}{|c|c|c|c|c|c|c|}
\hline
\multirow{2}{*}{\textit{education}} & \multicolumn{3}{c|}{\textit{sex=male}} & \multicolumn{3}{c|}{\textit{sex=female}} \\ \cline{2-7} 
                           & \textit{Weighting}   &\textit{Regression}        &\textit{SCM}          & \textit{Weighting}      &\textit{Regression}              & \textit{SCM}          \\ \hline
0                          & 0.196      &0.086      & 0.164        & 0.048          &0.026           & 0.057        \\ \hline
1                          & 0.269      &0.214      & 0.239        & 0.066          &0.051           & 0.075        \\ \hline
2                          & 0.513      &0.491      & 0.498        & 0.211          &0.190           & 0.221        \\ \hline
3                          & 0.736      &0.781      & 0.741        & 0.416          &0.497           & 0.469        \\ \hline
4                          & 0.842      &0.933      & 0.859        & 0.485          &0.807           & 0.706        \\ \hline
\end{tabular}}
\caption{Expectation of the potential outcome for males and females in Adult dataset.}
\label{tab:system-level}
\end{table}

\subsubsection{Checking Equal Effort at the Group Level}
For the group level equality of effort, we split the Adult dataset into five groups by \textit{education}. Individuals with the same education value form one group. For each group, we calculate the expectations of potential outcome for males ($\mathbb{E}[Y_{D^+}(t)]$) and females  ($\mathbb{E}[Y_{D^-}(t)]$). Due to space limit, we only report in Table~\ref{tab:group-level} the expectations of the potential outcome variable for group one with \textit{education=0}. Each expectation is calculated using three methods. We can see the significant discrepancy between males and females in this group. We also observe the similar phenomena in other four groups.   When considering $\gamma = 0.5$, the minimum education value to achieve the outcome for males in this group is 3 (with all expectation values from three methods close to 0.7) whereas the minimum education level for females is 4. 

\begin{table}[]
\centering
\scalebox{0.7}{
\begin{tabular}{|c|c|c|c|c|c|c|}
\hline
\multirow{2}{*}{\textit{education}} & \multicolumn{3}{c|}{\textit{sex=male}} & \multicolumn{3}{c|}{\textit{sex=female}} \\ \cline{2-7} 
                           & \textit{Weighting}   &\textit{Regression}        & \textit{SCM}          & \textit{Weighting}      &\textit{Regression}              & \textit{SCM}          \\ \hline
1                          & 0.225      &0.232      & 0.227        & 0.071          &0.084           & 0.081        \\ \hline
2                          & 0.457      &0.462      & 0.467        & 0.205          &0.205           & 0.224        \\ \hline
3                          & 0.692      &0.694      & 0.719        & 0.418          &0.411           & 0.497        \\ \hline
4                          & 0.810      &0.870      & 0.842        & 0.497          &0.693           & 0.754        \\ \hline
\end{tabular}}
\caption{Expectations of the potential outcome for males and females with the original \textit{education}=0.}
\label{tab:group-level}
\end{table}

\subsubsection{Checking Equal Effort at the Individual Level}

To detect effort discrepancy at the individual level, we need to  first identify a subset of users $I$ with the same characteristics of the given individual and then split them into the male group ($I^+$) and female group ($I^-$).  We then calculate the expectations of potential outcome for the male group ($\mathbb{E}[Y_{I^+} (t)]$) and female group ($\mathbb{E}[Y_{I^+} (t)]$) with each treatment level $t$. Due to space limit, we only report in  Table~\ref{tab:individual-level} the results of three randomly chosen female users whose index numbers are 425, 9569, and 46437.  Both users 1 and 2 have the original education value 1 and user 3 has education value 0.    
As shown in Table ~\ref{tab:individual-level}, the expectations of outcome for $I^+$ are consistently higher than $I^-$, indicating the existence of discrimination in terms of equal effort for these three individuals. For example, results of user 3 show that the minimum effort  for her to achieve $0.5$-level outcome is education $t=4$ whereas the corresponding minimum effort to achieve the same level outcome is $t=3$ had she been a male.   

\begin{table}
\centering
\scalebox{0.65}{
\begin{tabular}{|c|c|c|c|c|c|c|}
\hline
\multirow{2}{*}{\textit{education}} & \multicolumn{2}{c|}{\textit{User 1}} & \multicolumn{2}{c|}{\textit{User 2}}& \multicolumn{2}{c|}{\textit{User 3}} \\ \cline{2-7} 
                           & \textit{sex=male}   &\textit{sex=female}        & \textit{sex=male}          &\textit{sex=female}     &\textit{sex=male}              & \textit{sex=female}         \\ \hline
0                          &        &           &           &        &0.012     & 0.006       \\ \hline
1                          &  0.022 & 0.007     &  0.058    &0.030   &0.051     & 0.024       \\ \hline
2                          &  0.085 & 0.036     &  0.206    &0.134   &0.188     & 0.096        \\ \hline
3                          &  0.282 & 0.159     &  0.523    &0.438   &0.501     & 0.317       \\ \hline
4                          &  0.624 & 0.487     &  0.823    &0.796   &0.813     & 0.669        \\ \hline
\end{tabular}}
\caption{Expectation of the potential outcome for three randomly chosen individuals.}
\label{tab:individual-level}
\end{table}

\subsection{Discrimination Removal}

We run our removal algorithm to remove discrimination in terms of equality of effort from the Adult dataset, and then run the discovery algorithm to further examine whether discrimination is truly removed in the modified dataset. For comparison, we include the removal algorithm (Denoted by DI) of \cite{feldman2015}, which removes discrimination from the demographic parity perspective. Basically, DI tries to modify $X$ such that the modified $\hat{X}$ cannot be used to predict $S$. The results show that, after executing our removal method (with $\lambda=5$), the average difference between $\mathbb{E}[Y_{D^+}(t)]$ and $\mathbb{E}[Y_{D^-}(t)]$ for all $t$s is $-0.0136$, indicating all effort discrepancy has been removed. However, the average difference for the DI algorithm is $0.2628$, showing that DI does not remove effort discrepancy. Regarding data utility loss in terms of $\chi^{2}$, our method also outperforms the DI algorithm in that the utility loss of our method is $34778$, while the utility loss of the DI algorithm is $37997$.

\section{Conclusions and Future Work}

In this paper, we proposed a new causal-based fairness notion called the equality of effort. 
Although previous fair notions can be used to judge discrimination from various perspectives (e.g., demographic parity, equal opportunity), they cannot quantify the (difference in) efforts that individuals need to make in order to achieve certain outcome levels. Our proposed notion, on the other hand, can help answer counterfactual questions like ``how much credit score an applicant should improve such that the probability of her loan application approval is above a threshold'', and judge discrimination from the equal-effort perspective.
To quantify the average effort discrepancy, we developed a general method under certain assumptions, as well as specific methods based on three widely used causal inference techniques. When equality effort is not achieved by a dataset, we also developed an optimization-based method to remove discrimination. 
In the experiments, we show that the Adult dataset does contain effort discrepancy at system, group, and also individual levels, and our removal method can ensure that the newly generated dataset satisfies equality of effort.

We made several assumptions in our paper including the no-hidden-confounder assumption, monotonicity of the expectation of outcome variable, and invertibility of outcome function. We also assumed one binary protected attribute and one binary decision for simplicity's sake. The no-hidden-confounder assumption is a common assumption for causal inference \cite{pearl2009causality} and widely adopted by causal inference based fair learning. The monotonicity assumption reflects the real world phenomena (the more effort, the better outcome). The invertibility assumption is used in our general method of calculating the average effort discrepancy without deriving the closed form of the inverse function. When this invertibility assumption is not held, we have presented in our algorithm (Lines 12-19) several inference methods that could also have their limitations. Moreover, we implicitly assumed that the discrimination detection algorithm knows the same information as the decision-maker, i.e., there are no omitted variables used in decision making but invisible to the discrimination detection. In our future work, we will study how to achieve equal effort fairness when some of those assumptions are not met in practice.  

In our paper, we used the change of treatment variable value as the effort needed to achieve a certain level of outcome and did not consider the real monetary or resource cost behind that change that are often not included in the data. If they are included in the data, the discrimination caused by these factors is known as indirect discrimination. We will study the use of path-specific effect/mediator analysis \cite{DBLP:conf/ijcai/ZhangWW17,nabi2018fair} to explicitly quantify the effect of treatment on final outcomes via proxy attributes. 

\section{ Acknowledgments}
This work was supported in part by NSF 1646654, 1920920, and 1940093.


\begin{thebibliography}{}

\bibitem[\protect\citeauthoryear{Burgette, Griffin, and
  McCaffrey}{2017}]{burgette2017propensity}
Burgette, L.; Griffin, B.~A.; and McCaffrey, D.
\newblock 2017.
\newblock Propensity scores for multiple treatments: A tutorial for the mnps
  function in the twang package.
\newblock {\em R package. Rand Corporation}.

\bibitem[\protect\citeauthoryear{Chiappa and Gillam}{2019}]{Chiappa2018Path}
Chiappa, S., and Gillam, T.~P.
\newblock 2019.
\newblock Path-specific counterfactual fairness.
\newblock In {\em AAAI'19}.

\bibitem[\protect\citeauthoryear{Dwork \bgroup et al\mbox.\egroup
  }{2012}]{dwork2012fairness}
Dwork, C.; Hardt, M.; Pitassi, T.; Reingold, O.; and Zemel, R.
\newblock 2012.
\newblock Fairness through awareness.
\newblock In {\em Proceedings of the 3rd Innovations in Theoretical Computer
  Science Conference},  214--226.
\newblock ACM.

\bibitem[\protect\citeauthoryear{Feldman \bgroup et al\mbox.\egroup
  }{2015}]{feldman2015}
Feldman, M.; Friedler, S.~A.; Moeller, J.; Scheidegger, C.; and
  Venkatasubramanian, S.
\newblock 2015.
\newblock Certifying and {{Removing Disparate Impact}}.
\newblock In {\em Proceedings of the 21th {{ACM SIGKDD International
  Conference}} on {{Knowledge Discovery}} and {{Data Mining}} - {{KDD}} '15},
  259--268.
\newblock {ACM Press}.

\bibitem[\protect\citeauthoryear{Hardt \bgroup et al\mbox.\egroup
  }{2016}]{Hardt2016Equality}
Hardt, M.; Price, E.; Srebro, N.; et~al.
\newblock 2016.
\newblock Equality of opportunity in supervised learning.
\newblock In {\em Advances in neural information processing systems},
  3315--3323.

\bibitem[\protect\citeauthoryear{Heidari, Nanda, and
  Gummadi}{2019}]{DBLP:journals/corr/abs-1903-01209}
Heidari, H.; Nanda, V.; and Gummadi, K.~P.
\newblock 2019.
\newblock On the long-term impact of algorithmic decision policies: Effort
  unfairness and feature segregation through social learning.
\newblock {\em CoRR} abs/1903.01209.

\bibitem[\protect\citeauthoryear{Holland}{1986}]{holland1986statistics}
Holland, P.~W.
\newblock 1986.
\newblock Statistics and causal inference.
\newblock {\em Journal of the American statistical Association}
  81(396):945--960.

\bibitem[\protect\citeauthoryear{Imbens}{2000}]{imbens2000role}
Imbens, G.~W.
\newblock 2000.
\newblock The role of the propensity score in estimating dose-response
  functions.
\newblock {\em Biometrika} 87(3):706--710.

\bibitem[\protect\citeauthoryear{Kilbertus \bgroup et al\mbox.\egroup
  }{2017}]{kilbertus2017avoiding}
Kilbertus, N.; Carulla, M.~R.; Parascandolo, G.; Hardt, M.; Janzing, D.; and
  Sch{\"o}lkopf, B.
\newblock 2017.
\newblock Avoiding discrimination through causal reasoning.
\newblock In {\em Advances in Neural Information Processing Systems},
  656--666.

\bibitem[\protect\citeauthoryear{Kusner \bgroup et al\mbox.\egroup
  }{2017}]{kusner2017counterfactual}
Kusner, M.~J.; Loftus, J.; Russell, C.; and Silva, R.
\newblock 2017.
\newblock Counterfactual fairness.
\newblock In {\em Advances in Neural Information Processing Systems},
  4066--4076.

\bibitem[\protect\citeauthoryear{Lichman}{2013}]{lichman2013}
Lichman, M.
\newblock 2013.
\newblock {{UCI Machine Learning Repository}}.
\newblock http://archive.ics.uci.edu/ml.

\bibitem[\protect\citeauthoryear{Luong, Ruggieri, and
  Turini}{2011}]{luong2011k}
Luong, B.~T.; Ruggieri, S.; and Turini, F.
\newblock 2011.
\newblock k-nn as an implementation of situation testing for discrimination
  discovery and prevention.
\newblock In {\em Proceedings of the 17th ACM SIGKDD international conference
  on Knowledge discovery and data mining},  502--510.
\newblock ACM.

\bibitem[\protect\citeauthoryear{McCaffrey \bgroup et al\mbox.\egroup
  }{2013}]{mccaffrey2013tutorial}
McCaffrey, D.~F.; Griffin, B.~A.; Almirall, D.; Slaughter, M.~E.; Ramchand, R.;
  and Burgette, L.~F.
\newblock 2013.
\newblock A tutorial on propensity score estimation for multiple treatments
  using generalized boosted models.
\newblock {\em Statistics in medicine} 32(19):3388--3414.

\bibitem[\protect\citeauthoryear{Nabi and Shpitser}{2018}]{nabi2018fair}
Nabi, R., and Shpitser, I.
\newblock 2018.
\newblock Fair inference on outcomes.
\newblock In {\em Proceedings of AAAI'18}, volume 2018.

\bibitem[\protect\citeauthoryear{Paszke \bgroup et al\mbox.\egroup
  }{2017}]{paszke2017automatic}
Paszke, A.; Gross, S.; Chintala, S.; Chanan, G.; Yang, E.; DeVito, Z.; Lin, Z.;
  Desmaison, A.; Antiga, L.; and Lerer, A.
\newblock 2017.
\newblock Automatic differentiation in pytorch.

\bibitem[\protect\citeauthoryear{Pearl}{2009}]{pearl2009causality}
Pearl, J.
\newblock 2009.
\newblock {\em Causality}.
\newblock Cambridge university press.

\bibitem[\protect\citeauthoryear{Romei and
  Ruggieri}{2014}]{romei2014multidisciplinary}
Romei, A., and Ruggieri, S.
\newblock 2014.
\newblock A multidisciplinary survey on discrimination analysis.
\newblock {\em The Knowledge Engineering Review} 29(05):582--638.

\bibitem[\protect\citeauthoryear{Rosenbaum and
  Rubin}{1983}]{rosenbaum1983central}
Rosenbaum, P.~R., and Rubin, D.~B.
\newblock 1983.
\newblock The central role of the propensity score in observational studies for
  causal effects.
\newblock {\em Biometrika} 70(1):41--55.

\bibitem[\protect\citeauthoryear{Russell \bgroup et al\mbox.\egroup
  }{2017}]{russell2017worlds}
Russell, C.; Kusner, M.~J.; Loftus, J.; and Silva, R.
\newblock 2017.
\newblock When worlds collide: integrating different counterfactual assumptions
  in fairness.
\newblock In {\em Advances in Neural Information Processing Systems},
  6414--6423.

\bibitem[\protect\citeauthoryear{Scheines \bgroup et al\mbox.\egroup
  }{1998}]{scheines1998tetrad}
Scheines, R.; Spirtes, P.; Glymour, C.; Meek, C.; and Richardson, T.
\newblock 1998.
\newblock The tetrad project: Constraint based aids to causal model
  specification.
\newblock {\em Multivariate Behavioral Research} 33(1):65--117.

\bibitem[\protect\citeauthoryear{Spirtes, Glymour, and
  Scheines}{2000}]{spirtes2000causation}
Spirtes, P.; Glymour, C.~N.; and Scheines, R.
\newblock 2000.
\newblock {\em Causation, prediction, and search}, volume~81.
\newblock MIT press.

\bibitem[\protect\citeauthoryear{Verma and Rubin}{2018}]{verma2018fairness}
Verma, S., and Rubin, J.
\newblock 2018.
\newblock Fairness definitions explained.
\newblock In {\em 2018 IEEE/ACM International Workshop on Software Fairness
  (FairWare)},  1--7.
\newblock IEEE.

\bibitem[\protect\citeauthoryear{Wu \bgroup et al\mbox.\egroup
  }{2019}]{DBLP:journals/corr/abs-1910-12586}
Wu, Y.; Zhang, L.; Wu, X.; and Tong, H.
\newblock 2019.
\newblock {PC}-fairness: {A} unified framework for measuring causality-based
  fairness.
\newblock {\em CoRR} abs/1910.12586.

\bibitem[\protect\citeauthoryear{Zafar \bgroup et al\mbox.\egroup
  }{2017}]{zafar2017fairness}
Zafar, M.~B.; Valera, I.; Rodriguez, M.~G.; and Gummadi, K.~P.
\newblock 2017.
\newblock Fairness constraints: Mechanisms for fair classification.
\newblock In {\em AISTATS}.

\bibitem[\protect\citeauthoryear{Zhang and
  Bareinboim}{2018a}]{zhang2018equality}
Zhang, J., and Bareinboim, E.
\newblock 2018a.
\newblock Equality of opportunity in classification: A causal approach.
\newblock In {\em Advances in Neural Information Processing Systems},
  3671--3681.

\bibitem[\protect\citeauthoryear{Zhang and
  Bareinboim}{2018b}]{zhang2018fairness}
Zhang, J., and Bareinboim, E.
\newblock 2018b.
\newblock Fairness in decision-makingâ€”the causal explanation formula.
\newblock In {\em Thirty-Second AAAI Conference on Artificial Intelligence}.

\bibitem[\protect\citeauthoryear{Zhang, Wu, and Wu}{2016}]{zhang2016b}
Zhang, L.; Wu, Y.; and Wu, X.
\newblock 2016.
\newblock Situation {{Testing}}-{{Based Discrimination Discovery}}: {{A Causal
  Inference Approach}}.
\newblock In {\em Proceedings of the {{Twenty}}-{{Fifth International Joint
  Conference}} on {{Artificial Intelligence}}, {{IJCAI}} 2016, {{New York}},
  {{NY}}, {{USA}}, 9-15 {{July}} 2016}, volume 2016-Janua,  2718--2724.
\newblock {IJCAI/AAAI Press}.

\bibitem[\protect\citeauthoryear{Zhang, Wu, and Wu}{2017a}]{zhang2017a}
Zhang, L.; Wu, Y.; and Wu, X.
\newblock 2017a.
\newblock Achieving {{Non}}-{{Discrimination}} in {{Data Release}}.
\newblock In {\em Proceedings of the 23rd {{ACM SIGKDD International
  Conference}} on {{Knowledge Discovery}} and {{Data Mining}}, {{Halifax}},
  {{NS}}, {{Canada}}, {{August}} 13 - 17, 2017},  1335--1344.
\newblock New York, New York, USA: {ACM Press}.

\bibitem[\protect\citeauthoryear{Zhang, Wu, and
  Wu}{2017b}]{DBLP:conf/ijcai/ZhangWW17}
Zhang, L.; Wu, Y.; and Wu, X.
\newblock 2017b.
\newblock A causal framework for discovering and removing direct and indirect
  discrimination.
\newblock In {\em Proceedings of the Twenty-Sixth International Joint
  Conference on Artificial Intelligence, {IJCAI} 2017},  3929--3935.

\bibitem[\protect\citeauthoryear{{\v{Z}}liobaite, Kamiran, and
  Calders}{2011}]{vzliobaite2011handling}
{\v{Z}}liobaite, I.; Kamiran, F.; and Calders, T.
\newblock 2011.
\newblock Handling conditional discrimination.
\newblock In {\em Data Mining (ICDM), 2011 IEEE 11th International Conference
  on},  992--1001.
\newblock IEEE.

\end{thebibliography}

\end{document}